%% file: main.tex
\documentclass[aps,amsmath,amssymb,superscriptaddress,twocolumn,showpacs,12pts, floatfix, showpacs,intlimits]{revtex4-1}

\usepackage[english]{babel}
\usepackage[utf8]{inputenc}
\usepackage{float}
\input{preamble}
\begin{document}
\title{Non-mechanical steering of the optical beam in spectral-domain Optical Coherence Tomography}

\author{Daniel F. Urrego}
\email[Correspondence email address: ]{daniel.urrego@icfo.eu}
\affiliation{ICFO -- Institut de Ciencies Fotoniques, The Barcelona Institute of Science and Technology, 08860 Castelldefels, Barcelona, Spain.\looseness=-1}
\author{Gerard J. Machado}
\affiliation{Department of Physics, Imperial College London, Prince Consort Road, SW7 2AZ London, United Kingdom.}
\author{Juan P. Torres}
\affiliation{ICFO -- Institut de Ciencies Fotoniques, The Barcelona Institute of Science and Technology, 08860 Castelldefels, Barcelona, Spain.\looseness=-1}
\affiliation{Department of Signal Theory and Communications, Universitat Politecnica de Catalunya, 08034 Barcelona, Spain\looseness=-1}

\date{\today} 

\begin{abstract}
We demonstrate in a {\it proof-of-concept} experiment spectral-domain optical coherence tomography where steering of the optical beam that probes the sample in a transverse scan does not make use of any mechanical element. Steering is done with the help of a spatial light modulator, that introduces a spatially-dependent phase between the two orthogonal polarization components of an optical beam, and some optical elements that control the polarization of light. We demonstrate that making use of the non-mechanical steering system considered here, we can reproduce the main traits of imaging with standard OCT that makes use of mechanical-assisted optical beam steering. 
\end{abstract}


\maketitle

\section{Introduction}
Optical coherence tomography (OCT) is a three-dimensional imaging technique  introduced in 1991 that makes use of a Michelson interferometer, where light in one arm illuminates the sample and light in the other arm serves as a reference \cite{Huang1991,Dresel1992}. The axial and transverse resolutions of OCT are independent.
To obtain information in the transverse direction (plane perpendicular to the beam propagation), OCT focuses light to a small spot that is scanned over the sample. To obtain information in
the axial direction (along the beam propagation), OCT uses a source of light with short coherence length that allows optical sectioning of the sample.
OCT is nowadays an active area of research as well as a mature technnology \cite{OCTbook2006,OCTbook2015,OCTbook2019} that finds applications in many areas of science and technology, from medicine, particularly ophthalmology \cite{drexler2008,everyday2006,Haeusler1998} to art conservation studies \cite{adler2007,liang2007}. 

The first OCT schemes used {\it time-domain OCT}, where axial sectioning of the sample was done by moving a mirror in the reference arm with the help of a fast-moving mechanical platform, such as a Galvo scanner or a 2D translation stage. The translation of the mirror in the reference arm severely limited the imaging speed achievable. Stratus OCT, introduced in 2002, achieved speeds of
$\sim 400$ A-scans per second \cite{octdevelopment2016}. In 1995, {\it Spectral} or {\it Fourier-domain} OCT was a major breakthrough for the development and rapid expansion of OCT \cite{Fercher1995}. In spectral-domain OCT, axial sectioning of the sample is done measuring the spectrum of the light back-scattered from the sample with a spectrometer, eliminating the need to move the position of the mirror in the reference arm. OCT can also be done with a laser that changes fast its frequency ({\it swept-source OCT}) \cite{chinn1997}, and both spectral-domain and swept-source OCT schemes provides a better axial resolution than time-domain OCT \cite{chomaIzatt2003,Leitgeb2003}. One advantage of swept-source OCT is that a spectrometer is not required, only a broadband detector. Spectral-domain OCT is currently the standard for ophthalmic instruments, with imaging speeds in excess of $\sim 25,000$ A-scans per second \cite{octdevelopment2016}.

In order to obtain an image of the sample with good transverse resolution, all the OCT schemes considered still require the use of a fast-moving mechanical device to scan a focused spot of light across the sample. We propose a significant innovation, demonstrated through a proof-of-concept experiment, which introduces a method for performing transverse scans of the sample without requiring a mechanical platform. The mechanical element is replaced by a spatial light modulator (SLM) in combination with some simple optical elements to modify the polarization of light \cite{moreno2007}. We will show that by engineering the phase introduced by groups of pixels of the SLM, one can perform the transverse scan of the light beam without any mechanical scan needed.

One word of caution might be appropriate here. We put forward a scheme, that in combination with spectral-domain OCT, allows to perform OCT fully in the optical domain without the need of any mechanical steering of the optical beam that probes the sample. Since we use a slow-response spatial light modulator (Hamamatsu X10468), our proof-of-principle demonstration does not achieve the scan speeds of state-of-the-art OCT systems. In spite of this, it can help usher in the development of alternative OCT schemes with potentially faster scan speeds than current OCT systems by avoiding the use of any kind of mechanical platforms.  

Non-mechanical laser beam steering, for applications such as beam steering in free-space laser communications, has been proposed and demonstrated using liquid crystal optical phased arrays \cite{nonmechanical2018}. Most commonly used liquid crystal devices share that they show a slow response, due to the slow rotation of the liquid crystal molecule. However, there is on-going research aimed at speeding up agile beam steering, developing new materials and arquitectures that allows faster change of molecules orientation \cite{fast2004,fast2013,fast2015}. Recently, a spatial light modulator architecture has been proposed that can achieve two-dimensional phase-only modulation at high speed in excess of GHz.  It makes use of a tunable two-dimensional array of vertically oriented, one-sided microcavities
that are tuned through an electro-optic material such as barium titanate \cite{fast2019}.

\begin{widetext}
\begin{minipage}{0.9\linewidth}
\begin{figure}[H]
\centering
\includegraphics[width=\linewidth]{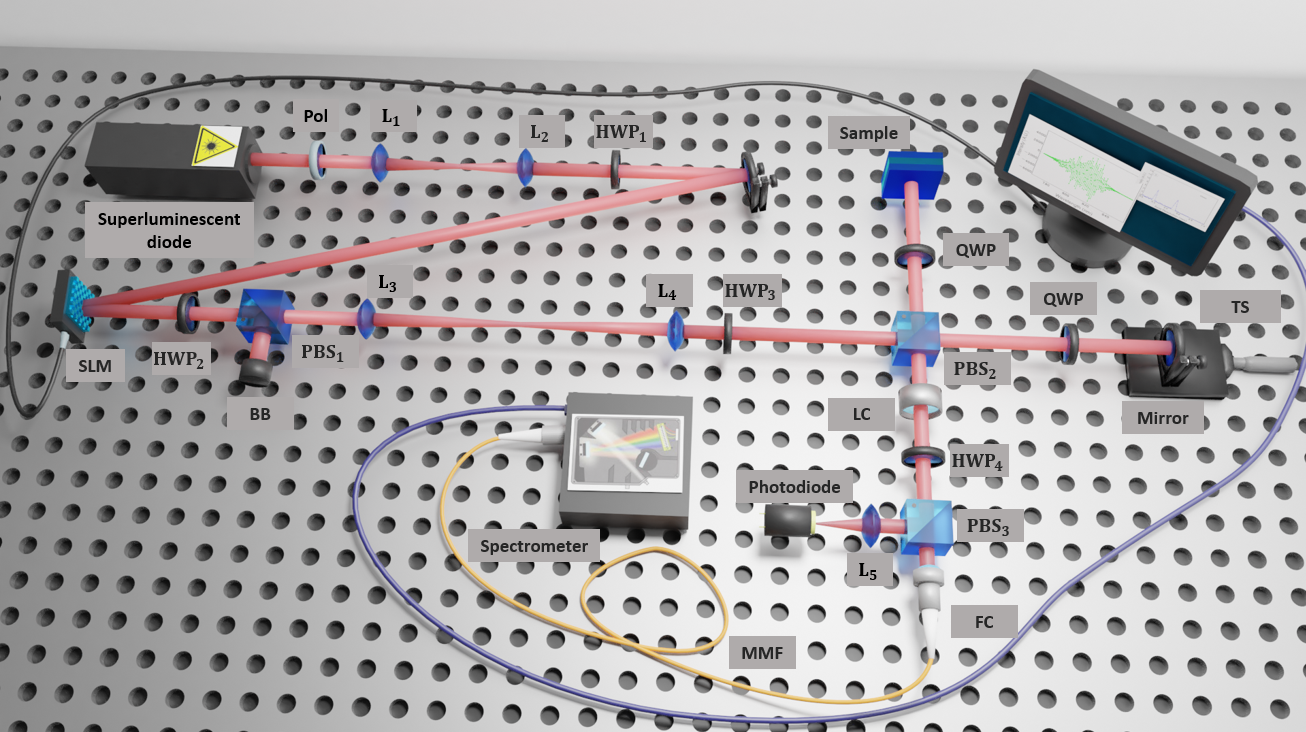}
\caption{Experimental setup for non-mechanical transverse steering of an optical beam in optical coherence tomography. A superluminescent diode illuminates an SLM that is the key element of the non-mechanical scanning system. The SLM introduces a different phase on a pixel-by-pixel basis. The output beam, with a new spatial shape, enters a polarization-sensitive Michelson interferometer.  A Liquid Crystal Variable Retarder introduces a controllable phase difference between the light beams coming from the sample and reference arms, that are recombined at the outport port of the Michelson interferometer.  Two interferograms are acquired employing a spectrometer with a phase difference  $\Delta \phi=0,\pi$. Several methods (see main text) are used to filter out the meaningless peaks of the Fourier transform signal, and enhance the sought-after peaks that carry information about the longitudinal structure of the sample.}
\label{fig: experimentalsetup}
\end{figure}
\end{minipage}
\end{widetext}

\section{Theory and Experimental setup}
Figure~\ref{fig: experimentalsetup} shows a detailed overview of the experimental setup. It consist of a polarization-sensitive Michelson interferometer, in combination with a spatial light modulator (SLM) and some optical elements that control the polarization of light.  A superluminescent diode (SLD) generates light with central wavelength  $810~\text{nm}$ and a bandwidth of $\sim 29.5~\text{nm}$. The polarization of the light generated by the SLD is selected to be horizontal with the help of a polarizer. The  spatial shape of the light beam is Gaussian, and its beam width is magnified by a telescope with lenses $\text{L}_1$ and  $\text{L}_2$, reaching a beam waist of $\text{w}_0=1.26~\text{mm}$. This is the light beam that will enter the non-mechanical transverse steering system of the OCT configuration.

\subsection{Non-mechanical steering of the optical beam}
We aim at generating a Gaussian-like beam with a small cross section in one transverse direction, that can be steered in the transverse plane, without the use of any mechanical platform, and still cover the whole area of the sample to be scanned. For achieving this goal we employ a half-wave plate ($\text{HWP}_1$), a spatial light modulator (SLM, Hamamatsu X10468, 1060 $\times$ 970 pixels), a second half-wave plate ($\text{HWP}_2$) and a polarizing beam splitter ($\text{PBS}_1$). Half-wave plate $\text{HWP}_1$ is rotated an angle $\theta$, so it generates a light beam with polarization 
\begin{equation}
\hat{e}_{in}=\cos 2\theta\, \hat{ \bf e}_H+\sin 2\theta\, \hat{\bf e}_V,
\label{polarization_in}
\end{equation}
where $\hat{ \bf e}_H$ and $\hat{ \bf e}_V$ designate horizontal and vertical polarization, respectively. 

Each pixel, or group of pixels, of the SLM introduce a spatially-dependent phase $\Phi(x,y)$ only to the vertical polarization component, so the polarization of the light beam after reflection from the SLM is
 \begin{equation}
  \hat{e}_{SLM}(x,y)=r_H\,cos 2\theta\,\hat{e}_\text{H} + r_V\,\sin 2\theta\,\exp \left[ i\Phi(x,y) \right] \hat{e}_\text{V},
\end{equation}
where we have considered that the reflection coefficients for the horizontal and vertical polarizations ($r_\text{H}$ and $r_\text{V}$) can be different. 

Half-wave plate $\text{HWP}_2$ is rotated an angle $\pi/8$ so that the polarization of the optical beam just before PBS$_1$ is  
\begin{eqnarray}
&  &  \hat{e}_{out}(x,y) =\frac{1}{\sqrt{2}} \Big\{ r_H \cos 2\theta+ r_V \sin 2\theta\, \exp \left[ i \Phi(x,y) \right] \Big\} \hat{\bf e}_H \nonumber \\ 
 & & + \frac{1}{\sqrt{2}} \left\{ r_H \cos 2\theta- r_V \sin 2\theta\, \exp \left[ i \Phi(x,y) \right]\right\} \hat{\bf e}_v. 
   \label{polarization_out}
\end{eqnarray}
We can choose the spatial dependence of $\Phi(x,y)$ so as to tailor the polarization of the light reflected by groups of pixels. In the ideal case of equal reflectivities ($r_H=r_V$), we can see from Eq.(\ref{polarization_out}) that for $\theta=\pi/8$, a phase $\Phi(x,y)=0$ would generate a reflected beam with horizontal polarization. PBS$_1$ will transmit this section of the beam with horizontal polarization, that will illuminate the sample in the optical coherence tomography scheme. For all other pixels, we can choose a phase $\Phi(x,y)=\pi$, so that this section of the light beam shows vertical polarization, which is reflected by PBS$_1$ and consequently discarded. 

However, the reflection coefficients in the SLM for both polarizations are different. We measured the values $r_\text{H}=0.91$ and $r_\text{V}=0.7$. We can tune the orientation angle $\theta$ of HWP$_1$ to minimize the intensity associated with the unwanted polarization components of the output beam, i.e., $I_\text{V}(\Phi=0)$, that is equivalent to minimizing $I_\text{H}(\Phi=\pi)$.  The intensity of the horizontal component of the output beam, before PBS$_1$ is
\begin{equation}
    I_\text{H}(x,y)=\frac{1}{2}\,\Big| r_H\,\cos 2\theta+r_V\,\sin 2\theta\, \exp \left[ i \Phi(x,y) \right] \Big|^2.
\end{equation}
The maximum and minimum values of the intensity of the horizontal polarization component are obtained for $\Phi(x,y)=0,\pi$, so that 
\begin{widetext}
\begin{minipage}{0.9\linewidth}
\begin{equation}
I_\text{H}^{max,min}(x,y)=\frac{1}{2}\Big[ |r_H|^2\,\cos^2 2\theta+|r_V|^2\,\sin^2 2\theta \pm 2\,|r_H|\, |r_V|\,\cos 2\theta\,\sin 2\theta \Big],
\end{equation}
\end{minipage}
\end{widetext}
where the positive sign corresponds to the phase $\Phi(x,y)=0$ and the negative sign to $\Phi(x,y)=\pi$.  The visibility $V=[I_\text{H}^{max}-I_\text{H}^{min}]/[I_\text{H}^{max}+I_\text{H}^{min}]$ is
\begin{equation}
    V=\frac{2|r_H|\,|r_V|\, \cos 2\theta\, \sin 2\theta}{|r_H|^2\,\cos^2 2\theta +|r_V|^2\, \sin^2 2\theta}.
\end{equation}
Taking in account the experimental values of $r_\text{H}=0.91$ and $r_\text{V}=0.70$, we get the maximum visibility for $\theta \approx 26.2^{\circ}$. This is the value of $\theta$ that we will use in the experiments.

\begin{figure}[t!]
\centering
\includegraphics[width=\linewidth]{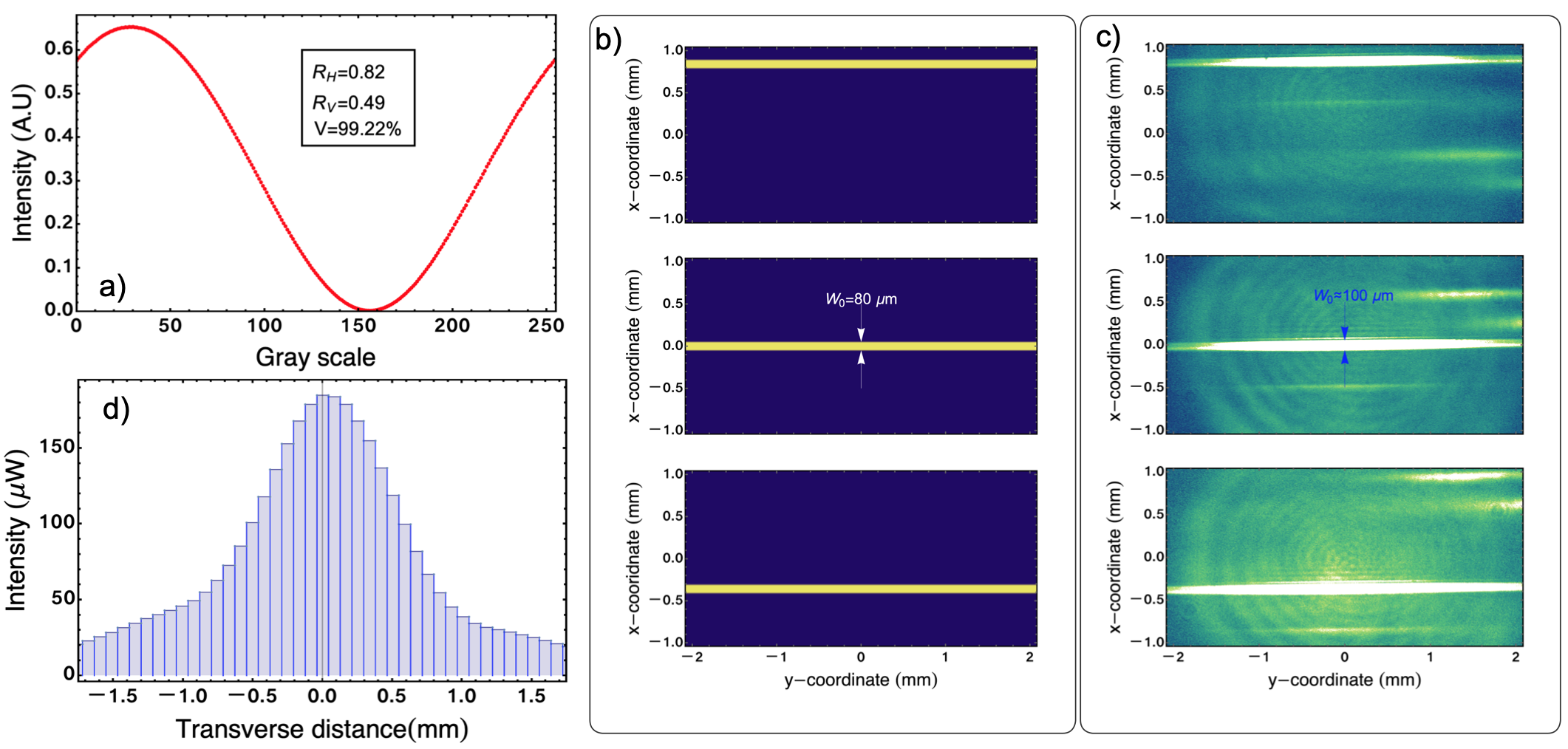}\\
\caption{Characteristics of the non-mechanical steering system. (a) Intensity of the output light as a function of the grayscale displayed in the SLM. The intensity is measured after $\text{PBS}_1$. The graph indicates the intensity change induced by the SLM. The maximum normalized standard error obtained in the measurement is $\sigma_{E}=6.6\times10^{-5}$. (b) Three different configurations of the phases introduced by groups of pixels on the screen of the SLM. The phase introduced by each pixel is selected using the corresponding value of the grayscale. (c) Pictures taken in the sample plane using the configuration depicted in (b). (d) Intensity distribution that arrives at the sample plane as a function of the translation of the bar displayed on the screen of the SLM. The maximum normalized standard error obtained in the measurement is $\sigma_{E}=2.47\times10^{-2}$.}
\label{SLMCharacterization}
\end{figure}

The control parameter that allows to tune the phase introduced by each pixel of the SLM  is the grayscale displayed on each pixel of the SLM's screen. Figure~\ref{SLMCharacterization}(a) shows the intensity of the light beam after  $\text{PBS}_1$ as a function of the grayscale displayed on each pixel of the SLM's screen. From this data, one can obtain the values of grayscale that maximize and minimize the intensity transmitted through $\text{PBS}_1$. These values correspond to the introduction of phases $0$ and $\pi$, respectively, and they are designated as states on/off of the pixels of the SLM. In our experimental set-up, the values of grayscales are 31 and 158, respectively. Fig.~\ref{SLMCharacterization}(b) shows three different arrays of on/off of pixels. The blue regions correspond to pixels in state off, and the yellow regions correspond to the pixels in state on. For the sake of example, we consider the case of the pixels forming a bar. The bar measures $80~\mu\text{m}$ in width and has a length exceeding $15~\text{mm}$.

Fig.~\ref{SLMCharacterization}(c) shows the shape of the output light beam generated by the corresponding distribution of on/off pixels, displayed in Fig.~\ref{SLMCharacterization}(b). The spatial profile of the beam is measured employing a CCD camera in the sample's plane. The bar can be displaced by changing the sets of pixels in the screen that are on/off, as shown in Fig.~\ref{SLMCharacterization}(b). The shape and intensity of the light beam that illuminates the sample depends on the characteristics of the input beam that illuminates the SLM and the position of the pixels that constitute the bar in the SLM.  Fig.~\ref{SLMCharacterization}(d) shows the maximum possible intensity that can reach the sample when the scan is done employing a bar with a width of $80~\mu\text{m}$. It is important to clarify that the measured intensity is acquired when all the incoming light to the interferometer is directed toward the sample. The maximum intensity is reached when the bar is in the middle of the Gaussian input beam, and its value is $\text{I}_\text{max}=180~\mu\text{W}$. 

\subsection{The polarization-sensitive Michelson interferometer}
A polarization-sensitive Michelson interferometer is the core of the OCT configuration. A $4f-\text{system}$ with focal lengths $\text{L}_3=\text{L}_4=400~\text{mm}$, allows the diffraction-less propagation of the light beam from the plane of the SLM's to the sample's plane.  The polarizing beam splitter $\text{PBS}_2$ splits the input light into the reference and sample arms. $\text{HWP}_3$ changes the polarization of the light beam coming from PBS$_1$, which is horizontal, and is used to unbalance the intensities between both arms, since the object in the sample arm will show in general a low reflectivity.  In the reference arm, there is a quarter-wave plate ($\text{QWP}$) oriented at $45^\circ$ and a mirror mounted on a motorized translation stage. In the sample arm, there is a $\text{QWP}$ at $45^\circ$ and the reflective sample to be probed. The light beams coming from the two arms of the interferometer are recombined in PBS$_2$, and due to the presence of the two QWP's, the output beam leaves the interferometer through the output port. 

\begin{figure}[t!]
\centering
\includegraphics[width=\linewidth]{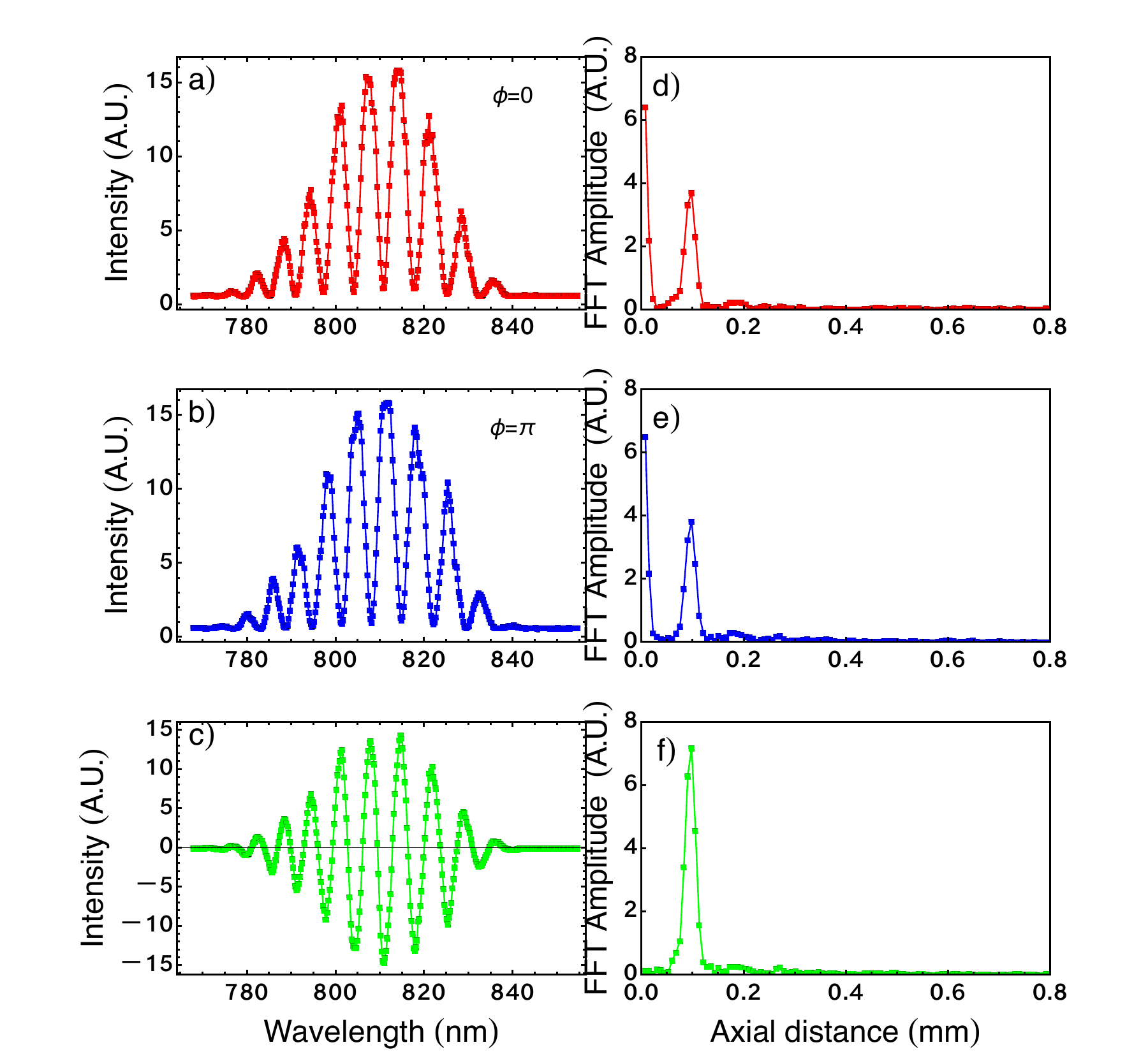}\\
\caption{(a) and (b): Spectrum measured with an spectrometer for two values of the phase introduced by the Liquid Crystal Variable Retarder ($\Phi=0$,$\pi$); (d) and (e): Signals obtained after Fourier-transform of the corresponding spectrum; (c) Signal resulting from subtracting the spectra shown in (a) and (b); (f) Signal obtained after Fourier-transform of the subtraction of the spectra shown in (c). }
\label{fig: SteppedPhase}
\end{figure}

The recombined beam traverses a Liquid Crystal Vaiable Retarder (LCVR) that adds an extra phase to the light coming from the reference arm (see Appendix A to see details of the relationship between the voltage and the phase introduced by the Liquid Crystal Variable Retarder). $\text{HPW}_4$ oriented at $\pi/8$ in combination with $\text{PBS}_3$ allows to separate the light beams coming from the two arms of the interferometer into diagonal and anti-diagonal orthogonal polarizations. Light transmitted by $\text{PBS}_3$ is sent to a spectrometer (Ocean Optics, HR4000) utilizing a multimode fiber (MMF), and the spectrum measured is analyzed and processed in order to retrieve the sought-after information of the sample. Light reflected from PBS$_3$ is sent to a photodiode (PD) with the help of a collecting lens with focal length $\text{L}_5=75~\text{mm}$.

\begin{figure}[t!]
\centering
\includegraphics[width=\linewidth]{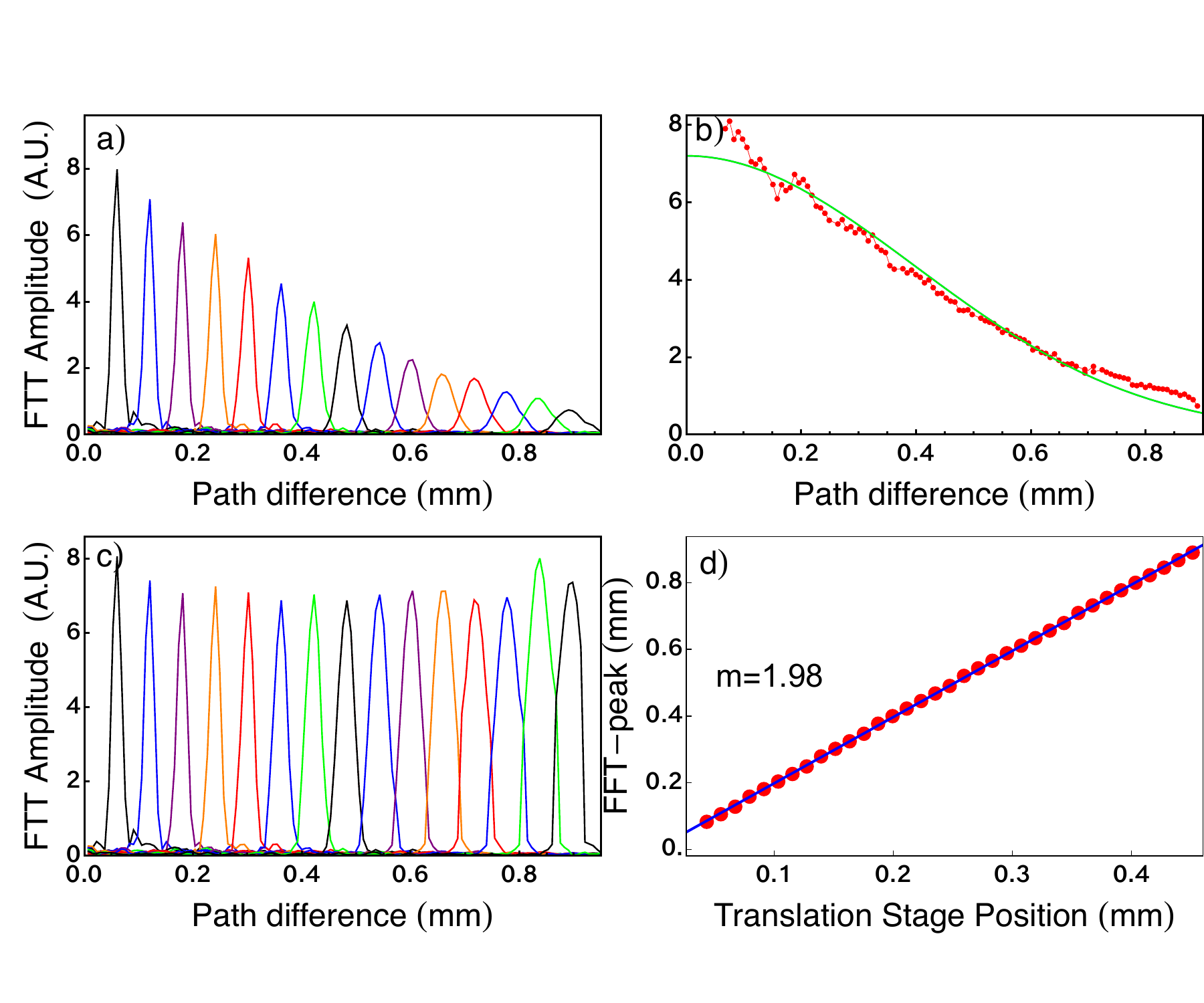}\\
\caption{Measurement of the location of a moving mirror, using spectral-domain OCT and applying post-processing corrections. (a): Intensity of the signal resulting from the Fourier-transform of the subtraction of two spectra with phase difference $\pi$, for some selected positions of the moving mirror. Every signal (different colors) corresponds to a position of the mirror along the reference arm. (b): The dots corresponds to the amplitudes of the peaks measured and shown in (a). The solid line is a fit to this data, and is the key element for doing the sensitivity decay correction in all experiments shown in this paper. (c): Signal obtained after applying the sensitivity decay correction. (d): Path length difference measured by Optical Coherence Tomography as function of the path length difference  induced by the motorized translation stage. The dots are the experimental data, and the solid line is a fit. The slope of the line matches the expected path length difference generated, $m_{\text{theo}}=2$.}
\label{fig: SensitivyDecay}
\end{figure}

\subsection{Stepped Phase shifting interferometry}
For any axial scan, i.e. for a specific transverse position, two spectra with a phase difference $\Delta \phi=\pi$, introduced by the Liquid Crystal Variable Retarder, are measured and subsequently subtracted. The resulting signal is Fourier transformed with the help of a standard Fast-Fourier transform (FFT) routine (see Appendix B for further details about how to perform the Fourier-transform of the signal measured with the spectrometer). Without subtraction,  the Fourier-transformed signal, contains a DC peak, and some auto-correlation peaks that appear due to interference of multiple reflections in the sample. The presence of these peaks make it difficult the characterization of the longitudinal structure of the sample.  By subtracting the two spectra with a phase difference of $\pi$,  all of these peaks are filtered out. The remaining peaks, that are enhanced, are the cross-correlations ones, the ones which bear relevant information about the structure of the sample   \cite{stepped2002}. 

\begin{figure}[t!]
\centering
\includegraphics[width=\linewidth]{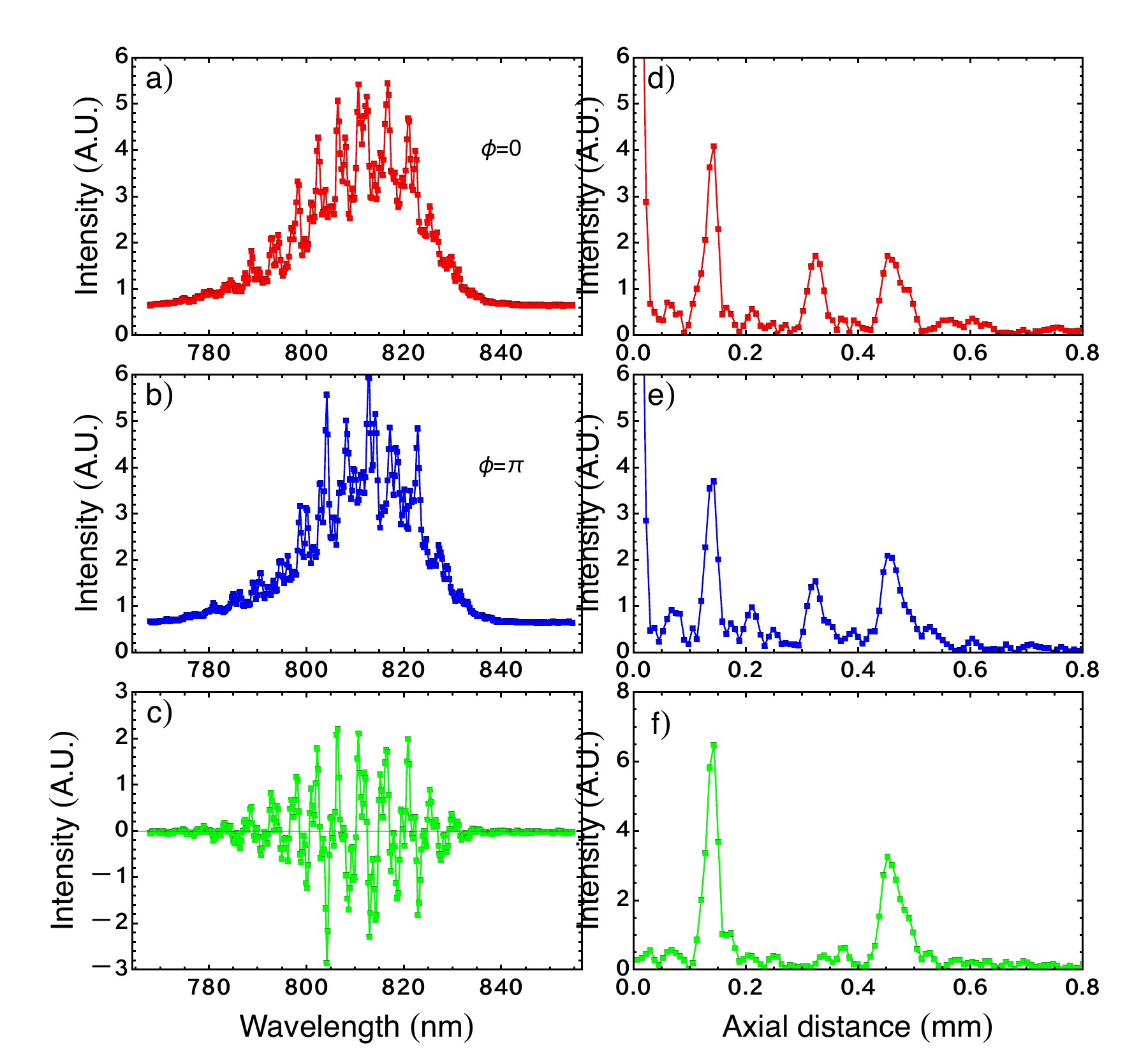}\\
\caption{(a) and (b): Spectrum measured with an spectrometer for two values of the phase introduced by the Liquid Crystal Variable Retarder ($\Phi=0$,$\pi$); (d) and (e): corresponding signals obtained after Fourier-transform of the spectrum; (c) Signal resulting from subtracting the spectra shown in (a) and (b); (f) Signal obtained after Fourier-transform of the subtraction of spectra shown in (c). Dots: experimental data. Solid lines: best fit to the experimental data.}
\label{fig: SteppedPhase2}
\end{figure}

\section{Demonstration of Optical Coherence Tomography with a non-mechanical steering system}
\subsection{First example: location of the position of a single-layer sample}
The Fourier-domain OCT apparatus is first characterized by employing a mirror instead of a multi-layered sample. This is equivalent to considering a single-layer sample with a reflection coefficient of about $\sim 0.97$. For the sake of convenience, the characterization is made by changing the position of the mirror in the reference arm. The motorized translation stage controls the path difference between the two arms of the interferometer. 

Figures~\ref{fig: SteppedPhase}(a) and~\ref{fig: SteppedPhase}(b) show an example of the experimental spectra obtained for the case of a single-layer sample (mirror) with $\phi=0$ and $\phi=\pi$, respectively. Figures~\ref{fig: SteppedPhase}(d) and ~\ref{fig: SteppedPhase}(e)  show the signal obtained after Fourier transform of the spectrum measured and shown in Figs.~\ref{fig: SteppedPhase}(a) and (b). One can see the presence of a strong DC peak. Figure~\ref{fig: SteppedPhase}(c) is the result of the subtraction of signals shown in Figs.~\ref{fig: SteppedPhase}(a) and (b), and Fig.~\ref{fig: SteppedPhase}(f) is the corresponding Fourier transform. The DC peak has been filtered out, and the only remaining peak allows to retrieve the position of the mirror.

Figure~\ref{fig: SensitivyDecay}(a) shows the signal obtained after Fourier-transform of the subtraction of the two spectra with phase difference $\pi$, for some selected positions of the mirror. Each peak of the Fourier Transform correspond to a position of the moving mirror. One would expect to observe the same peak amplitude for all positions of the mirror, since the reflectivity of the object does not change, only its location. However, for increasing path length differences between light reflected from the two arms of the Michelson interferometer, the amplitude of the peaks of the signal diminishes. This decay is due to the finite pixel size of the detector in the spectrometer \cite{decay2003,decay2004,decay2007}. To correct for this effect in all experiments, we will apply the sensitivity decay correction depicted in Fig.~\ref{fig: SensitivyDecay}(b). After this, we obtain Fig.~\ref{fig: SensitivyDecay}(c).  Considering the design of the Michelson interferometer, varying the position of the sample results in a twofold increase in the path length difference. Figure~\ref{fig: SensitivyDecay}(d) shows the positions of the peaks of the signal as function of the sample's position, which is controlled by a motorized translation stage. The solid line is a linear fit. The slope of the linear fit is $m_{exp}=1.98$, and it agrees with the theoretically expected value of $m_{theo}=2$.

\subsection{Second example: Three-dimensional image of a multilayer sample}
\begin{figure}[t!]
\centering
\includegraphics[width=\linewidth]{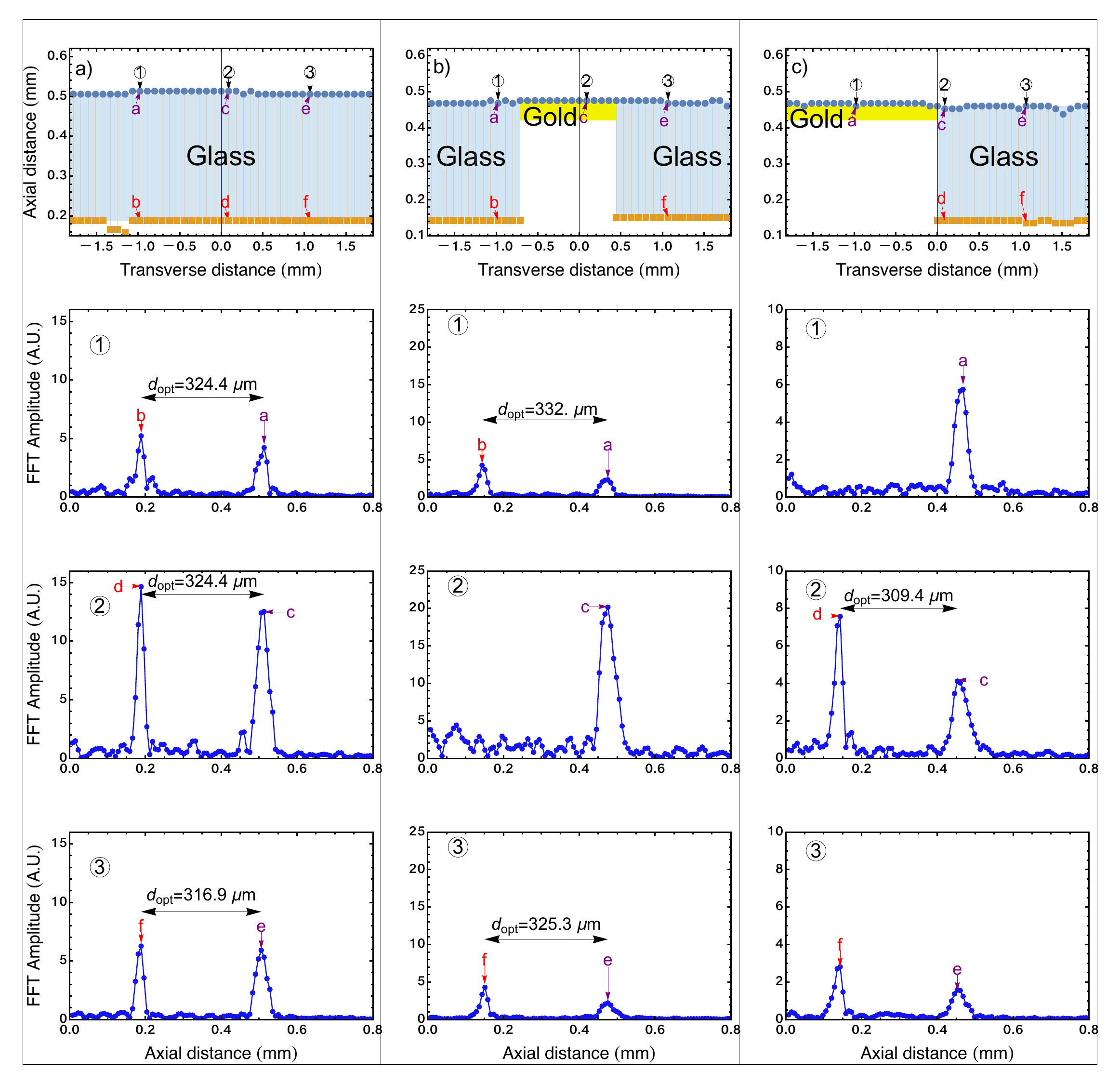}\\
\caption{OCT images of the samples. The axial scan is achieved by using Fourier-domain OCT, and the transverse scan is done employing the non-mechanical steering system described above. The samples shown in (b) and (c) consist of a piece of glass with a certain area covered with a thin film of gold, working as a mirror. For each sample, the three figures below show the signal obtained for three different transverse positions along the sample. The Fourier transform peaks indicate the positions of various layers within the sample.}
\label{fig: Samples}
\end{figure}

In order to prove the feasibility of the OCT scheme with a non-mechanical steering system implemented here, we do OCT with three different samples (see Fig.~\ref{fig: Samples}): (1) a piece of glass (refractive index $n_{\text{glass}}=1.5$); (2) a piece of glass with a film of gold of thickness $40~\text{nm}$ that covers the middle of the surface; and (3) a piece of glass with a fringe of gold of thickness $40~\text{nm}$ and width $1~\text{mm}$ that covers half of the surface of the glass. The gold film works as a mirror and the glass as a two-layer sample. 

Figures~\ref{fig: SteppedPhase2}(a) and~\ref{fig: SteppedPhase2}(b) show an example of the experimental spectra obtained for a bi-layer sample (a piece of glass with no gold on the surface) with $\phi=0$ and $\phi=\pi$, respectively. Figure~\ref{fig: SteppedPhase2}(d) and~\ref{fig: SteppedPhase2}(e) show the signal resulting of the Fourier transform of the corresponding spectrum. Figure~\ref{fig: SteppedPhase2}(c) is the result of subtracting the two spectra with phase difference of $\pi$, and Fig.~\ref{fig: SteppedPhase2}(f) is the signal after Fourier transform. We can see that the DC term, and the autocorrelation terms present in Figs.~\ref{fig: SteppedPhase2}(d) and (e) have been filtered out. 

Figures~\ref{fig: Samples}(a), (b) and (c) shows the experimental results obtained after doing the transverse scans in the three samples using the light beam with the bar shape shown in Fig.~\ref{SLMCharacterization}(b). Figures~\ref{fig: Samples}(a), (b) and (c) corresponds to the three samples described above. The numbers in each image designate a specific position of the transverse scan. The letters in each image designate a specific longitudinal position in the sample. For the transverse locations where there is no gold on top of the piece of glass, all insets show two peaks given that we are observing a two-layer sample. The distance between the peaks allows us to estimate the optical thickness of the piece of glass. In the transverse locations where there is gold, we observe a single peak that corresponds to a single-layer sample (a mirror). 

\section{Conclusions}
We have demonstrated, in a proof-of-principle experiment, spectral-domain Optical Coherence Tomography where beam steering of the optical beam to perform transverse scans of the sample is done without using any mechanical element. Instead we steer the optical beam with the help of a spatial light modulator and some optical elements that control the polarization of light.

The standard SLMs that we use in our proof-oc-concept experiments can change the phase map displayed by arrays of pixels with speeds of around $60$-$200$ Hz. Although our proof-of-principle demonstration does not achieve the scan speeds of state-of-the-art OCT systems, it can help usher in the development of alternative OCT schemes with potentially faster scan speeds than current OCT systems by avoiding the use of any kind of mechanical platforms.  Recent proposals for phase-only two-dimensional modulators show that one can achieve high-speed modulations in excess of a few GHz, that is one order of magnitude faster that the speeds achieved nowadays with fast mechanical devices.

In 2006, clinical OCT instruments scanned at a rate of $\sim 400$ axial scans per second \cite{OCTbook2006,octdevelopment2016}. Nowadays, most spectral-domain OCT systems can make axial scan at rates of tens of KHz  \cite{octdevelopment2016,Indian2017}. The scheme proposed here, in combination with the development of faster phase-only two-dimensional modulators can contribute to continue the increase of the rate axial scans in OCT. Moreover, the scheme demonstrated here can also be combined with the use of Digital Micromirror Devices (DMD), that can reach pattern rates of $\sim 32$ KHz, comparable to current scan rates for spectral domain OCT.

\section*{Acknowledgements}
This work is part of the R$\&$D project CEX2019-000910-S, funded by the Ministry of Science and innovation (MCIN/ AEI/10.13039/501100011033/). It has also been funded by Fundació Cellex, Fundació Mir-Puig, and from Generalitat de Catalunya through the CERCA program. We acknowledge suport from the project 20FUN02 “POLight”, that has received funding from the EMPIR programme co-financed  by the  Participating  States  and  from  the  European  Union's  Horizon  2020  research  and innovation programme. W also acknowledge financial support from project QUISPAMOL (PID2020-112670GB-I00) funded by MCIN/AEI /10.13039/501100011033. GJM was supported by the Secretaria d’Universitats i Recerca del Departament d’Empresa i Coneixement de la Generalitat de Catalunya and European Social Fund (FEDER).

\bibliography{References}

\appendix
\section{Characterization of the Liquid Crystal Variable Retarder}
A liquid crystal variable retarder (LCVR) is a cell filled with nematic liquid crystal molecules with ordered orientation. A differential voltage applied to the cell produces a tilt in the molecules' orientation. The effect of this tilt is the introduction of a phase in a polarization component, i.e., $\vec{E}_\text{H}+\vec{E}_\text{V} \Longrightarrow \vec{E}_\text{H}+\exp (i\varphi)\,\vec{E}_\text{V}$.

\begin{figure}[h!]
\centering
\includegraphics[width=\linewidth]{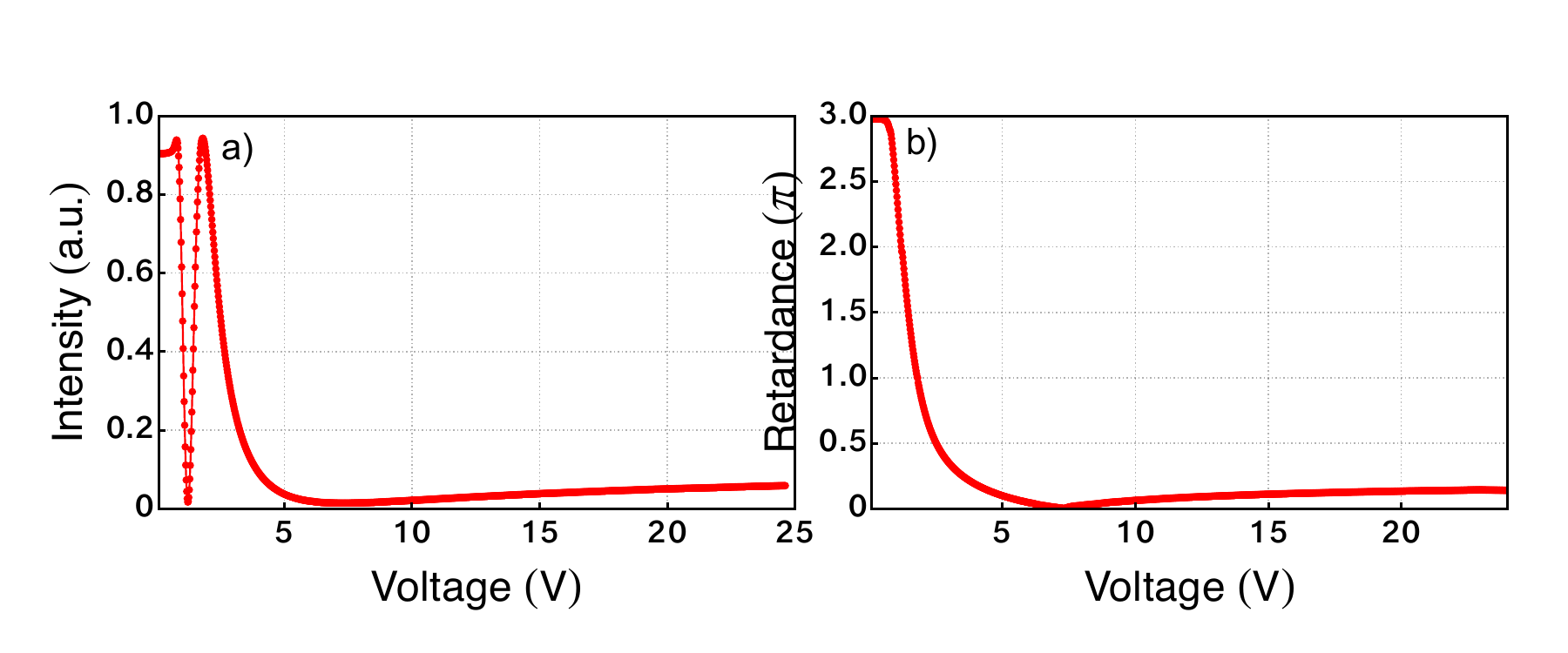}\\
\caption{Characterization of the liquid crystal variable retarder. (a) Intensity of the polarization component of the electric field as a function of the differential voltage. Each point in the graph corresponds two the average of 1000 samples. The voltage is varied in steps of $0.02~\text{mV}$. The maximum standard error obtained is $\sigma_{E}=9.6\times10^{-4}$. (b) Retardance (or phase) recovered from the intensity curve.}
\label{fig: LCR_Characterization}
\end{figure}

The characterization of the LCVR is performed by setting an elliptical polarization to the input electric field and observing the intensity of horizontal or vertical polarization when the induced voltage is changed. Fig. \ref{fig: LCR_Characterization}(a) shows the intensity of the horizontal component of the electric field as function of the voltage applied. The intensity $I$ is 
\begin{equation}
I=\frac{I_{max}}{2} \Big[ 1-\cos \varphi \Big],
\end{equation}
where $I_{max}$ is the maximum intensity~\cite{LopezTellez2014}.  As a result, the phase introduced by LCVR can be obtained as 
\begin{equation}
\varphi=\cos^{-1} \Big[ 1-\frac{2I}{I_{max}} \Big].
\end{equation} 
It is necessary to perform the unwrapping method to guarantee the phase continuity. Fig. \ref{fig: LCR_Characterization}(b) shows the phase (or retardance) as a function of the voltage applied. 

\section{Calculation of the signal resulting from the Fourier transform of the spectrum measured in an axial scann of Fourier-domain OCT}
The spectrum $S(\lambda)$ is measured with a spectrometer. The spectrum is rewritten as function of the wavenumber $k=2 \pi/\lambda$. Considering the Jacobian of the transformation, the k-spectrum $S(k)$ is
\begin{equation}
S(k) = \frac{2\pi}{k^2} S(\lambda)
\end{equation}
The spectrum is re-sampled to obtain a function $S(k)$ with equally-spaced $k$-values. The Fourier transform of the re-sampled spectrum is calculated as

\begin{equation}
{\cal F} [S(k)] =
\int dk\, S(k) \,\exp \left( ikz \right)
\end{equation}
where $z$ is the axial position along the sample.

\end{document}

%% file: preamble.tex
\usepackage{amsthm}
\usepackage{mathtools}
\usepackage{physics}
\usepackage{xcolor}
\usepackage{graphicx}
\usepackage[left=23mm,right=13mm,top=35mm,columnsep=15pt]{geometry} 
\usepackage{adjustbox}
\usepackage{placeins}
\usepackage[T1]{fontenc}
\usepackage{lipsum}
\usepackage{csquotes}